\documentclass[twocolumn]{svjour3}          
\smartqed  
\usepackage{amsmath}
\usepackage{graphicx}
\usepackage{epstopdf}
\begin{document}

\title{Geometrical Localization Algorithm for 3-D Wireless Sensor Networks}
\author{Rajesh Kumar   \and
        Sushil Kumar\and
        Diksha Shukla\and
        Ram Shringar Raw 
}
\institute{(1,2,3) Jawaharlal Nehru University, New Delhi   \at
              \email{rajeshjnu2006@gmail.com}\\
              \email{skdohare@gmail.com} \\
              \email{diksha.jnu08@gmail.com} \\
           \and
           (4) Ambedkar Institute of Advanced Communication Technologies \& Research, Delhi, India \at
           \email{rsrao08@yahoo.in}
}
\date{Accepted and Published by  Journal of Wireless Personal Communication, Springer : June, 2014 \\
The final version of publication is available at link.springer.com\\
Link: http://link.springer.com/article/10.1007\%2Fs11277-014-1852-6 }
\maketitle
\begin{abstract}
In this paper, we propose an efficient range free localization scheme for large scale three dimensional wireless sensor networks. Our system environment consists of two type of sensors, randomly deployed static sensors and global positioning system equipped moving sensors. These moving anchors travels across the network field and broadcast their current locations on specified intervals. As soon as the sensors which are deployed in random fashion receives three beacon messages (known locations broadcasted by anchors), they computes their locations automatically by using our proposed algorithm.
One of our significant contributions is, we use only three different beacon messages to localize one sensor, while in the best of our knowledge, all previously proposed methods use at least four different known locations. The ability of our method to localize by using only three known locations not only saves computation, time, energy, but also reduces the number of anchors needed to be deployed and more importantly reduces the communication overheads. Experimental results demonstrate that our proposed scheme improves the overall efficiency of localization process significantly.
\end{abstract}
\section{Introduction} Wireless Sensor Networks are composed of large numbers of sensors which are equipped with limited memory, computing, sensing capabilities and power source on-board \cite{AkyIF,AReview}. Recent advancements in wireless communications and electronics have enabled the development of low-cost, low-power and multi-functional sensors that are small in size and communicate in short distances. These cheap, smart, wireless enabled sensors which can be deployed in large numbers , and provide unprecedented opportunities for various kind of applications. Some of the important modern application of wireless sensor networks are military surveillance, environmental monitoring, habitat monitoring and structural monitoring etc. \cite{Applications1,Applications2,Applications3,Applications4}. \\
Many of the aforementioned applications require large scale wireless sensors network and the locations of these sensors to be known. Deploying huge number of sensors at the fixed known locations is a difficult task and sometimes practically impossible. One of the solutions to this problem is to deploy sensors in random fashion and computing their locations on the basis of few known locations.
Numerous schemes have been proposed to compute the location of sensors and all of them have their advantages and disadvantages. \\To evaluate localization
schemes, researchers have reported various performance metrics. Some of these performance metrics are localization accuracy, computational complexity, energy efficiency, time taken, number of anchors to be deployed, and communication overheads. We have designed an algorithm which uses
only three beacon messages (known locations) to compute the location of a sensor. Which in turn reduces the required number of anchors to be deployed to localize
the whole network. Moreover reduces the required communications and frequency of the broadcast of the beacon messages by moving sensors and hence reduces the energy consumption and communication overhead.
Consequently, overall process of the localization process is improved significantly.
\subsection{Problem Definition} Localization is a process of computing locations of the sensors which are randomly deployed in a wireless sensor network. Mathematically, our system environment of wireless sensor network can be modeled as multi-hop network and can be represented by a graph G = (V, E) where V is set of sensors which in turn
is a combination of two different sets say U and A and E is the wireless communication path among them. The set U consists of sensors which are randomly
deployed and static in nature. Whereas, set A consists of moving sensors equipped with GPS and hence they know their location. GPS equipped these sensors
are also known as anchors who broadcast their locations on specified intervals. We assume that all these sensors have homogeneous communication range R.
Let locations of be represented as $\{X_{a}, Y_{a}, Z_{a}\}$ and locations of static sensors of set U be represented as $\{X_{u}, Y_{u}, Z_{u}\}$
which is unknown in our case.\\
One of the easiest hardware solution is to have GPS equipped sensors and deploy them in the network field. However, due to the large scale nature of the
wireless sensor networks, and the cost of GPS makes this approach impractical. Therefore, researchers propose a variety of software solutions to localize
the sensors using some known locations or utilizing other sensors capability. Most of the previous localizations schemes can be categorized either into range-based
or range-free. Both of these have their advantages and disadvantages.
\subsection{Previous Methods} 
\begin{figure}
    \centering
    \includegraphics[width=3.5in]{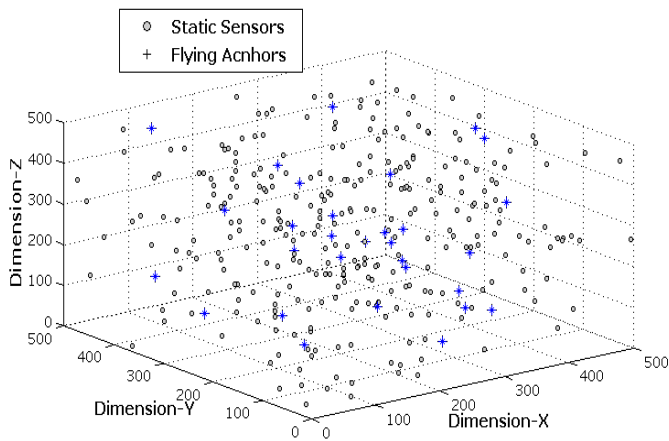}
    \caption{Three Dimensional Wireless Sensor Network Environment Having Flying Anchors (10\% of total number of sensors)  }
    \label{SystemEnvironment}
\end{figure}

A series of range-based schemes uses Time of Arrival (ToA)\cite{TOABased}, Time Difference of Arrival (TDoA)\cite{TDOABased}, Angle of Arrival (AoA)\cite{AOABased}, Received Signal Strength Indicator (RSSI)\cite{RSSIBased} etc. to compute the location of the sensor. Various other scheme use
Multidimensional Scaling (MDS) \cite{MDSBased}, Radio Interferometric Measurement(RIM) \cite{RIMBased}, DV-distance\cite{DVDistanceBased},
DV-hop\cite{DVHopeBased}, etc to localize. These range-based techniques are fairly accurate but are computationally expensive and require highly expensive special equipments.\\
Apart from aforementioned range based methods researchers have proposed range-free techniques in which position of a node is computed on the basis of
information transmitted by nearby anchor nodes or neighboring (already localized) nodes. The range-free schemes are not as accurate as range-based but are
useful in many of the modern applications high localization accuracy is not desired. Moreover, the performance of recently proposed range-free schemes
is comparable to that of the range based. In addition, these schemes are far cheaper than the range-based techniques\cite{RangeFreeAdvantages}.
A series of range-free schemes have been proposed and some of them uses APIT \cite{RangeFree}, moving anchor\cite{ChinesePaper} , three dimensional multilateration approach \cite{MultiLaterationBased},centroid scheme \cite{NovelCentroid}, weighted centroid\cite{WCentroidBased} etc.\\
Range free localization schemes\cite{RangeFree1,RangeFree2,RangeFree3} basically based on two different methodologies. The first method is to deploy few
GPS equipped sensors (known locations) along with the a large number of non GPS equipped sensors (unknown locations) in a random fashion. Now, using the
known location of GPS equipped sensors, we compute the location of other sensors \cite{GPSEnabledThesis}. The second method is to deploy a large number of non
GPS equipped sensors and use few GPS equipped moving sensors (anchors). These anchors travel across the network field (see fig.\ref{SystemEnvironment}) and broadcast their current locations on specified interval. As soon the non GPS equipped sensor receives required number of beacon messages, it computes its own location using some algorithm \cite{ChinesePaper,MNNITPaper,AnchorBased,AnchorBased2}. First method is highly erroneous in some cases, for instance, error gets propagated from one sensor to another. While the second method which uses moving anchor concepts is practically tough but it gives better accuracy.\\
To the best of our knowledge most of the proposed schemes till date which follows the second method require at least four beacon messages (four known locations)
to compute the location of unlocated sensor in \textsl{three dimensional wireless sensor network}. Similarly, at least three beacon are required three
beacon messages to find the position of a node in \textsl{two dimensional wireless sensor network}. The more the number of beacon messages are used, the more they are expensive
in terms of computations, time, energy and communication overheads\cite{LimitedAnchorNode,PowerEfficiency}. Hence reducing the number of required beacon messages to localize a node will improve the performance of the whole process. This paper focuses on the same i.e. reducing the number of beacon messages(eventually \# of anchors ) required to localize nodes and successfully reduces this requirement by twenty five percent.
\subsection {Three Dimensional Wireless Sensor Network Environment}
A pictorial representation of three dimensional wireless sensor network in our system environment is given in fig.\ref{SystemEnvironment}. Star labeled points corresponds to the anchor nodes whereas
circle labeled points are randomly deployed sensors. These anchors travels through the whole sensing space broadcasting their current locations, other sensors will receive these locations (any three
non linear different locations (beacon messages)) and compute their locations themselves using the received location information.The number of static sensor nodes are usually large in number whereas the number of moving sensors (anchors) are much lesser. These
sensors are deployed in network field and are designed to execute some specific task.

The rest of the paper is organized as follows. We discuss the related work in section 2 and present the algorithmic design in detail in section 3.
We then present the simulations and results in section 4 and our conclusion in section 5.
\section{Related Work}
Previous works which are most closely related to this paper are \cite{ChinesePaper,MNNITPaper}. In these paper, authors have mentioned different methods for calculating the location of the
sensor networks following the anchor-based and range-free localization strategy.
Chia-Ho Ou et al. in \cite{ChinesePaper} proposes a solution for localization which works on basic geometric principle that, "a perpendicular line passing through the center of a sphere's circular cross section also passes through the center of that sphere". The algorithm proposed by this paper can be summarized as follows \\

\begin{figure}
    \centering
    \includegraphics[width=3.5in]{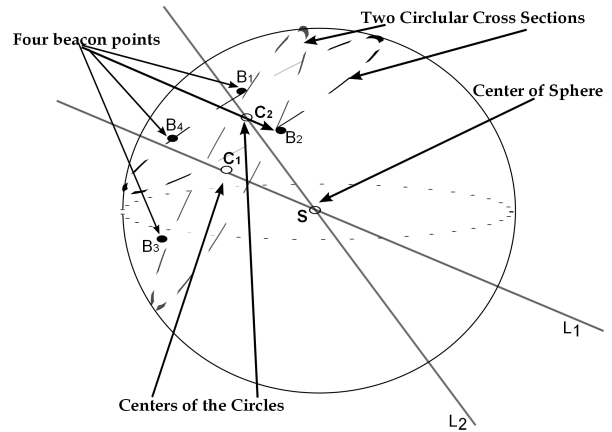}
    \caption{Two circular cross sections from four beacon points $B_{1}, B_{2} , B_{3} and B_{4}$. Two lines $L_{1}$ passing through center of circle $C_{1}$ and $L_{2}$ passing through center of $C_{2}$. Point of intersection of $L_{1}$ and $L_{2}$ is the sensors position S  }
    \label{TwoCircularcrossSection}
\end{figure}

1) Select any nonlinear \textit{four} beacon points as shown in the fig.\ref{TwoCircularcrossSection} by following the process explained in Sec 3.2.1 which is also illustrated through \ref{SelectionOfBeaconPoints}\\

2) Let (circles $C_1$ and $C_2$) be center of any two circular cross sections (circles) formed by these four points beacon points as shown in fig.\ref{TwoCircularcrossSection}\\

3) Find the center of these circles $C_{1},$ and $C_{2}$ \\

4) Now find the lines $L_{1}$ and $L_{2}$ Passing through the center of the circles $C_{1},$ and $C_{2}$ respectively and also perpendicular to the corresponding circular cross section as shown in the fig.\ref{TwoCircularcrossSection}.\\

5) Find the intersection point of the $L_{1}$ and $L_{2}$ which will definitely be the center of the sphere i.e. S and the desired location of the sensor node (see. fig.\ref{TwoCircularcrossSection}).\\

Vibha et al. \cite{MNNITPaper} also proposed the similar solution but have applied another basic geometric principle i.e. "if any point is at the surface of sphere then it will satisfy the sphere equation".
The algorithm proposed in this scheme can be summarized as follows\\

1) Select any nonlinear \textit{four} beacon points as shown in the fig.\ref{TwoCircularcrossSection} by following the process explained in Sec 3.2.1 which is also illustrated through \ref{SelectionOfBeaconPoints}\\

2) Substitute these four beacon points in equations of sphere to get four different equations (similar to \ref{FirstBeaconOnSphere}, \ref{SecondBeaconOnSphere} and \ref{ThirdBeaconOnSphere}.

3) The range for all the sensors are the same hence, by eliminating R from these equations, we get three different equations.\\

4) On solving these three equations, we get the desired center of sphere, which is the approximated position of the sensor.\\

We need at least four beacon points to apply any of the aforementioned algorithms. The sensor node which is to be localized will have to wait until it gets \textsl{at least four non-linear beacon points}(messages).
which delays the process, and results in more energy consumption. The number of beacon messages required by an algorithm is directly proportional to average localization time, beacon overhead,  communication overhead, energy consumption and number of anchors to be deployed etc. If we can somehow reduce the number of beacon messages required for localizing a node, it will reduce each and every parameter mentioned above. This is the motivation behind proposing the new scheme. Our proposed approach uses \textit{only three beacon points} to localize a particular node which itself a significant improvement because number of beacon messages required is directly proportional to the communication overhead, energy consumption and number of anchors to be deployed.

Therefore, if we are able to find the position of a node using only three beacons, we are reducing communication overhead, energy consumption and the number of anchors need to be deployed by twenty five percent. Moreover, number of computations and variables needed to compute the position of a node are also significantly reduced in our approach.

%

Chia-Ho Ou et al. in \cite{ChinesePaper} have considered the chord selection criteria vs points in a plane where chord selection criteria can be used to avoid beacons. This criterion selects those chords that are built with these beacons having angle \textgreater 10 degrees between them. Otherwise, it will lead to the location of the center of the sphere above or below the actual center. This problem will also be there for larger angle between them. But, in our approach occurrences of all coplanar beacons only lead to non determination of the center. Only one non-coplanar beacon will be sufficient to determine the center effectively. Hence, there is a low probability of discarding any position information obtained through beacon messages which is in turn will save time, computation and communication overheads.
\section{Proposed Work}
In this paper we have similar assumptions as that of \cite{MNNITPaper,ChinesePaper}. The assumptions are \\

1) Static nodes are the wireless sensor nodes which are deployed in random fashion and are need to be localized. \\

2) All static and dynamic sensor nodes have identical sensing ability, computational ability, ability to communicate and identical communication range R. The connectivity region of each node can be represented by a sphere of radius R, having the sensor node at its center. \\

3) Anchor nodes are GPS equipped, which moves across the network and broadcast their current location on specified intervals. These nodes are very few in number and are homogeneous in nature. \\

4) The communication range of mobile sensor nodes are assumed not to change drastically during the entire localization process. As soon as any static sensor node comes within its communication range, it will receive the broadcast message.\\

Moreover, our scheme uses both the geometrical theorems used in \cite{ChinesePaper} and \cite{MNNITPaper}. In addition to these theorems we are using few basic geometric theorems and vector calculus.
Geometrical theorems which are used to compute the location of the sensor in this paper are summarized below\\

1) A perpendicular line passing through the center of a sphere's circular cross section also passes through the center of that sphere. As shown in fig.\ref{FigureCircularcrossSection} the perpendicular line $L_{1}$ passes through both the center of the circular cross-section C and the center of the sphere S.\\

2) If any point is at the surface of sphere then it will satisfy the equation of sphere. For example standard equation of sphere having its center at $(X_{s}, Y_{s}, Z_{s})$ and radius R can be given as\\
\begin{equation}
\label{EquationOfSphere}
(X - X_s)^2 + (Y - Y_s)^2 + (Z - Z_s)^2 = R^2
\end{equation}
and If there is a point $(X_{1}, Y_{1}, Z_{1})$ which lies on the sphere then it will satisfy the equation of the sphere. Thus,
\begin{equation}
\label{PointLyingOnSphere}
(X_1 - X_s)^2 + (Y_1 - Y_s)^2 + (Z_1 - Z_s)^2 = R^2
\end{equation}

3) Area of a triangle ABC is the half of magnitude of cross product of side vectors. If ABC is a triangle then area of  this triangle can be given as,
\begin{equation}
\label{AreaOfTriangle}
\triangle ABC = \frac{1}{2}|\overrightarrow{BA}\times \overrightarrow{BC}|
\end{equation}

4) If two vectors are inclined at an angle then sum of these vectors is represented by the diagonal of the parallelogram formed by these two vectors. For example, if P and Q are two vectors inclined at some angle then
\begin{equation}
\label{ResultantOfPandQ}
\overrightarrow{P} +\overrightarrow{Q} = \overrightarrow{R}
\end{equation}
\subsection{Algorithmic Design}
Algorithmic design of our method is given below. In this algorithms, we are using only three beacon messages and compute the location of sensors using received messages\\

\begin{figure}
    \centering
    \includegraphics[width=3.5in]{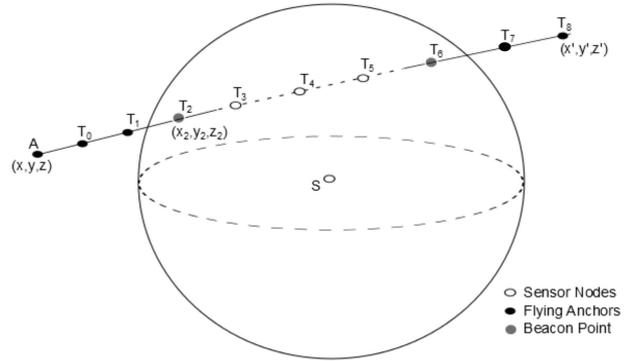}
    \caption{Process of selecting a beacon message: Flying Anchor A is moving from (X, Y, Z) to (X', Y', Z') and broadcasting the current location messages at an interval of time t, where $T_{n+1}-T_{n}$. S is the sensor location of which to be determined. Moreover, only two beacons are selected not three to maintain non-linearity. }
    \label{SelectionOfBeaconPoints}
\end{figure}

1) Select any nonlinear \textit{three} beacon points as shown in the fig.\ref{TwoCircularcrossSection} by following the process explained in Sec 3.2.1 which is also illustrated through \ref{SelectionOfBeaconPoints}\\

2) Substitute these three beacon points in the equation of sphere \ref{SphereEquation} to get three different equations \ref{FirstBeaconOnSphere}, \ref{SecondBeaconOnSphere} and \ref{ThirdBeaconOnSphere}.

3) Since, in our assumption the communication range is the same (R) for all the sensors, by eliminating R from equations \ref{FirstBeaconOnSphere}, \ref{SecondBeaconOnSphere} and \ref{ThirdBeaconOnSphere}, we get two equations
\ref{FirstRequiredEquation} and \ref{SecondRequiredEquation}\\

4) Consider one circular cross sections produced from joining all the three chosen beacons points $B_{1}$, $B_{2}$ and $B_{3}$ as illustrated in fig.\ref{FigureCircularcrossSection}.\\

\begin{figure}
    \centering
    \includegraphics[width=3.0in]{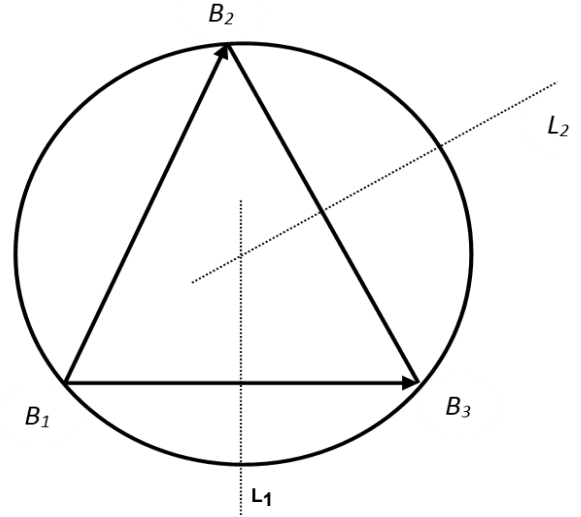}
    \caption{Corollary: Two perpendicular bisectors $L_{1}$ and $L_{2}$ are passing through the center of the circle}
    \label{CenterUsingPerpendicularBisectorMethod}
\end{figure}

5) Compute the center of the circular cross section (circle) $C_{1}$ either by using vector method described in \textsl{section 3.2.4 (i)} and illustrated in fig. \ref{CenterUsingVectorMethod} \textbf{or}
perpendicular bisector method described in \textsl{section 3.2.4 (ii)} and illustrated in fig. \ref{CenterUsingPerpendicularBisectorMethod} \\

6) Find the line equation of line $L_{1}$ i.e. equation.\ref{SphereCenterLieOnL1} passing through the center of the circle $C_{1}$ and perpendicular to the circular cross section as shown in fig. \ref{CenterUsingPerpendicularBisectorMethod}.\\

7) On solving the equations \ref{FirstRequiredEquation}, \ref{SecondRequiredEquation}, and \ref{SphereCenterLieOnL1}, we get the required location of the sensor.\\
These steps are discussed in detail in section 3.2.
\subsection{Computation of Location of the Sensors }
This section is to explain the step by step computations required in the proposed method.
\subsubsection{Process of Selecting a Beacon Message} The process of receiving or selecting a beacon messages (known location) which participates in computation are beautifully explained in \cite{ChinesePaper}. We also follow their method for selecting beacon messages, where each moving anchor periodically broadcasts beacon messages detailing its ID, position, and timestamp. Meanwhile, each sensor node maintains a set of beacon points and a Visitor List. The beacon point is regarded as an endpoint on the communication sphere of the sensor node, and the Visitor List maintains the IDs and respective lifetimes of the moving anchors currently passing through this sphere. \\
When a sensor node receives a beacon message from a moving anchor, it first checks whether the moving anchor is already recorded in its Visitor List. If it is missing, a beacon point, i.e., the current position of the moving anchor, is logged, and the ID and corresponding lifetime of the moving anchor are added to the Visitor List. If the moving anchor is already present in the Visitor List, the beacon message is ignored and the lifetime of the anchor is simply extended. When the lifetime of the moving anchor expires, its final beacon message is recorded as a beacon point and the corresponding entry for the anchor is deleted from the Visitor List.
For example, as illustrated in the fig.\ref{SelectionOfBeaconPoints} a moving anchor A (the anchors) moves from (X, Y, Z) to (X', Y', Z') broadcasting beacon messages at an interval of t, where t = $T_{i+1} - T_{i}$, and i = 0,1,2,3...8. The beacon messages at $T_{2}$ is considered as a beacon point $(A, (X_{2}, Y_{2}, Z_{2}))$ by sensor node S. Now node S adds an entry $(A, (X_{2}, Y_{2}, Z_{2}), T_{2} + L)$ to its visitor list, where L is the predefined lifetime of the moving anchor and has a value larger than the beacon interval t $(L = a \times t, a \textgreater 1$.\\
The lifetime of node A is increased by L when it arrives at $T_{3}, T_{4}, T_{5},$ and $T_{6},$ respectively. Once node 'A' moves out of the communication sphere of node S and its lifetime expire, its beacon message at $T_{6}$ is logged as a beacon point and node S deletes the entry for node A from its Visitor List. Similarly we proceed and select three beacons say $B_{1}, B_{2}$ and $B_{3}$.
\subsubsection{Getting Two Equations Using the Equation of The Sphere and The Three Beacon Points}
Consider, the first beacon message received by the static sensor node is broadcasted from the position \\$B_{1}$ $(X_{1}, Y_{1}, Z_{1})$ and that of the second and third are from $B_{2} (X_{2}, Y_{2},Z_{2})$ and $B_{3} (X_{3}, Y_{3}, Z_{3})$ respectively. Let S $(X_{s}, Y_{s}, Z_{s})$ be the sensor node whose location is to be determined and R be the communication range of all sensor nodes. Now the communication range of sensor node L can be given as:
\begin{equation}
\label{SphereEquation}
(X -X_s)^2 + (Y - Y_s)^2 + (Z - Z_s)^2 = R^2
\end{equation}
Beacons are selected within the spherical range of the sensor node, therefore coordinates of the beacons \\$B_{1} (X_{1}, Y_{1}, Z_{1}) , B_{2} (X_{2}, Y_{2},Z_{2}) and B_{3} (X_{3}, Y_{3}, Z_{3})$ will satisfy the equation of the sphere given in Equation(\ref{SphereEquation}) and it will give Equation(\ref{FirstBeaconOnSphere}), (\ref{SecondBeaconOnSphere}), and (\ref{ThirdBeaconOnSphere}) as below:
\begin{equation}
\label{FirstBeaconOnSphere}
(X_1 - X_s)^2 + (Y_1 - Y_s)^2 + (Z_1 - Z_s)^2 = R^2
\end{equation}
\begin{equation}
\label{SecondBeaconOnSphere}
(X_2 - X_s)^2 + (Y_2 - Y_s)^2 + (Z_2 - Z_s)^2 = R^2
\end{equation}
\begin{equation}
\label{ThirdBeaconOnSphere}
(X_3 - X_s)^2 + (Y_3 - Y_s)^2 + (Z_3 - Z_s)^2 = R^2
\end{equation}

We have assumed that the range 'R' of each sensor is identical, so from Equation(\ref{FirstBeaconOnSphere}), (\ref{SecondBeaconOnSphere}), and (\ref{ThirdBeaconOnSphere}) we get (\ref{FirstRequiredEquation}), (\ref{SecondRequiredEquation}) as below:
\begin{multline}
\label{FirstRequiredEquation}
(X_1 - X_s)^2 + (Y_1 - Y_s)^2 + (Z_1 - Z_s)^2 \\
= (X_2 - X_s)^2 + (Y_2 - Y_s)^2 + (Z_2 - Z_s)^2 \\
\end{multline}
\begin{multline}
\label{SecondRequiredEquation}
(X_1 - X_s)^2 + (Y_1 - Y_s)^2 + (Z_1 - Z_s)^2\\
= (X_3 - X_s)^2 + (Y_3 - Y_s)^2 + (Z_3 - Z_s)^2
\end{multline}

\subsubsection{Unique Circle Produced By Joining From Three Nonlinear Beacons Points}
\begin{figure}
    \centering
    \includegraphics[width=3.0in]{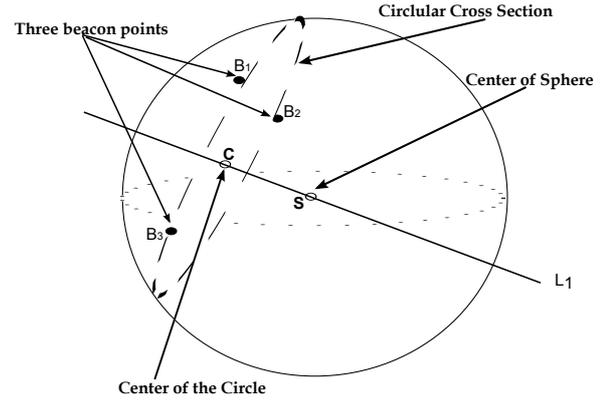}
    \caption{Unique circular cross sections produced from three beacons $B_{1}, B_{2}, B_{3}$ (selected as detailed in fig.\ref{SelectionOfBeaconPoints}) and having center C}
    \label{FigureCircularcrossSection}
\end{figure}

We know that a unique circle is produced using three nonlinear points, fig.\ref{FigureCircularcrossSection} illustrate this fact. There are three beacons $B_{1}, B_{2}, B_{3}$ which produces a circle with center $C_{1}$. Any line which passes through the center of the circle C will pass through the center of the sphere S in our case it is $L_{1}$.
\subsubsection{Computation of the Center of the Circle}
\begin{figure}
    \centering
    \includegraphics[width=3.0in]{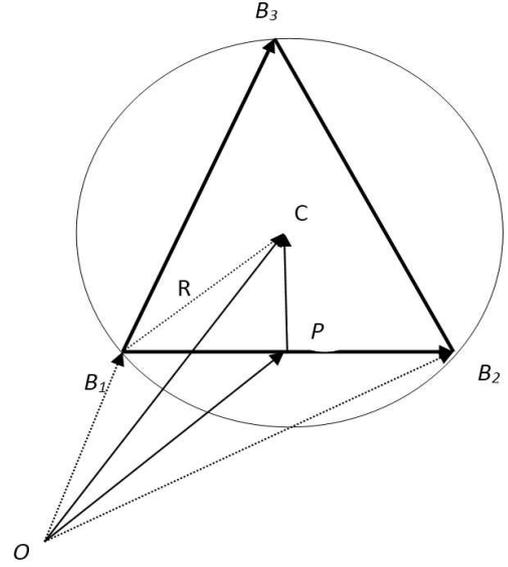}
    \caption{C is the center of the circle formed by joining beacon points $B_{1},B_{2},B_{3}$. We compute the center of the circle using the process detailed in Sec 3.2.4. vectors being used in the process are sketched here }
    \label{CenterUsingVectorMethod}
\end{figure}

We have proposed two different methods to compute coordinates of the center of the circle and named them Vector method and Perpendicular bisector method respectively.\\
\textbf{\textit{(i)	Vector method:}}
From the fig.\ref{CenterUsingVectorMethod}, we can see that
\begin{equation}
\label{CenterVectorEquation}
\overrightarrow{OC} =\overrightarrow{OP} + \overrightarrow{PC}
\end{equation}
\begin{equation}
\label{CenterVectorEquation2}
\overrightarrow{OC} = \frac {\overrightarrow{OB_1} + \overrightarrow{OB_2}} {2} + \overrightarrow{PC} = X_c \hat{i}+ Y_c \hat{j}+ Z_c \hat{k}
\end{equation}
Further, let C $(X_{c}, Y_{c}, Z_{c})$ is the center of the circular cross section, radius is R and the area of the triangle $(B_{1}B_{2}B_{3})$ formed by joining beacon points is a and PC can be given as
\begin{equation}
\label{eqn_example}
PC =\sqrt{R^2 - (a/2)^2 }
\end{equation}
where a, the area of the triangle $(B_{1}B_{2}B_{3})$ is
\begin{equation}
\label{eqn_example}
a =\sqrt{(X_2-X_1)^2+(Y_2-Y_1)^2+(Z_2-Z_1)^2}
\end{equation}
and R is the radius (sensing range of a sensor) can be computed as
\begin{equation}
\label{eqn_example}
R =\frac {abc}{4 \triangle}
\end{equation}
a, b, c sides of the triangle $(B_{1}B_{2}B_{3})$ can be computed as
\begin{equation}
\label{eqn_example}
a =\sqrt{(X_2-X_1)^2+(Y_2-Y_1)^2+(Z_2-Z_1)^2}
\end{equation}
\begin{equation}
\label{eqn_example}
b =\sqrt{(X_3-X_2)^2+(Y_3-Y_2)^2+(Z_3-Z_2)^2}
\end{equation}
\begin{equation}
\label{eqn_example}
c =\sqrt{(X_1-X_3)^2+(Y_1-Y_3)^2+(Z_1-Z_3)^2}
\end{equation}
\begin{equation}
\label{eqn_example}
\triangle = \frac{1}{2}|\overrightarrow{B_1B_2}\times \overrightarrow{B_1B_3}|
\end{equation}
\begin{equation}
\label{eqn_example}
\triangle = \frac{1}{2} \sqrt{\triangle _x ^2 + \triangle _y ^2 + \triangle _z ^2}
\end{equation}
\begin{equation}
\label{eqn_example}
\triangle _x = (Y_2-Y_1)(Z_3-Z_1)-(Z_2-Z_1)(Y_3-Y_1)
\end{equation}
\begin{equation}
\label{eqn_example}
\triangle _y = (Z_3-Z_1)(X_2-X_1)-(X_3-X_1)(Z_2-Z_1)
\end{equation}
\begin{equation}
\label{eqn_example}
\triangle _z = (X_2-X_1)(Y_3-Y_1)-(Y_2-Y_1)(X_3-X_1)
\end{equation}
The unit vector in the direction of PC can be given as
\begin{equation}
\label{eqn_example}
\hat{PC} = \frac{\overrightarrow{N} \times \overrightarrow{B_1B_2}}{|\overrightarrow{N} \times \overrightarrow{B_1B_2}|} = d_1 \hat{i}+ d_2 \hat{j}+ d_3 \hat{k}
\end{equation}
\begin{equation}
\label{eqn_example}
\overrightarrow{N}= \overrightarrow{B_1B_2} \times \overrightarrow{B_1B_3}= (N_x,N_y,N_z)
\end{equation}
\begin{equation}
\label{eqn_example}
N_x = (Y_2-Y_1)(Z_3-Z_1)-(Z_2-Z_1)(Y_3-Y_1)
\end{equation}
\begin{equation}
\label{eqn_example}
N_y = (Z_3-Z_1)(X_2-X_1)-(X_3-X_1)(Z_2-Z_1)
\end{equation}
\begin{equation}
\label{eqn_example}
N_z = (X_2-X_1)(Y_3-Y_1)-(Y_2-Y_1)(X_3-X_1)
\end{equation}
\begin{equation}
\label{CalculatingD1}
d_1 = \frac {N_y(Z_2-Z_1)-N_z(Y_2-Y_1)} {m}
\end{equation}
\begin{equation}
\label{CalculatingD2}
d_2 = \frac {N_z(X_2-X_1)-N_x(Z_2-Z_1)} {m}
\end{equation}
\begin{equation}
\label{CalculatingD3}
d_3 = \frac {N_x(Y_2-Y_1)-N_y(X_2-X_1)} {m}
\end{equation}
The value of m mentioned in the equations (\ref{CalculatingD1}),(\ref{CalculatingD2}) and (\ref{CalculatingD3}) can be computed using equation (\ref{ValueOfM}) given on the next page.
Finally, coordinates of the center $(X_{c}, Y_{c}, Z_{c})$ of the circular cross section
using equation(\ref{CenterVectorEquation2}), can be given as\\
\begin{equation}
\label{eqn_example}
X_c = \frac {X_1+X_2} {2}+ d_1 * PC
\end{equation}
\begin{equation}
\label{eqn_example}
Y_c = \frac {Y_1+Y_2} {2}+ d_2 * PC
\end{equation}
\begin{equation}
\label{eqn_example}
Z_c = \frac {Z_1+Z_2} {2}+ d_3 * PC
\end{equation}
\begin{figure*}[!t]
\normalsize
\begin{equation}
\label{ValueOfM}
m =\sqrt{{N_y(z_2-z_1)-N_z(y_2-y_1)}^2+{N_z(x_2-x_1)-N_x(z_2-z_1)}^2+ {N_x(y_2-y_1)-N_y(x_2-x_1)}^2}
\end{equation}
\end{figure*}
\begin{figure*}[!t]
\normalsize
\begin{equation}
\label{ComputationForT}
t=\frac{(X_1^2+Y_1^2+Z_1^2)-(X_2^2+Y_2^2+Z_2^2)-2(X_1-X_2)X_c-2(Y_1-Y_2)Y_c-2(Z_1-Z_2)Z_c}{2(X_1-X_2)N_X+2(Y_1-Y_2)N_Y+2(Z_1-Z_2)N_z}
\end{equation}
\end{figure*}
\textbf{\textit{(ii) Perpendicular bisector method: }}
Consider the circular cross section , the normal vector to the perpendicular bisector of the chord vector $B_{1}B_{3}$ i.e. $(L_{1})$ can be generated by the cross product of and vector $\overrightarrow{N}$ and $B_{1}B_{3}$ and similarly $L_{2}$ can be generated by the cross product of  and $\overrightarrow{N}$ and $B_{3}B_{2}$, i.e.
\begin{equation}
\label{eqn_example}
\overrightarrow{L_1} = \overrightarrow{N} \times \overrightarrow{B_1B_3} = (l,m,n)
\end{equation}
\begin{equation}
\label{eqn_example}
\overrightarrow{L_2}= \overrightarrow{N}\times\overrightarrow{ B_3B_2} = (p,q,r)
\end{equation}
Equations of the straight lines $L_{1} and L_{2}$ can also be written as follows:
\begin{equation}
\label{eqn_example}
L_1 : \frac {X-a_1}{l}=\frac {Y-b_1}{m}=\frac {Z-c_1}{n}=t_1
\end{equation}
\begin{equation}
\label{eqn_example}
L_2 : \frac {X-a_2}{p}=\frac {Y-b_2}{q}=\frac {Z-c_2}{r}=t_2
\end{equation}
Where$(a_{1},b_{1},c_{1})$ and $(a_{2},b_{2},c_{2})$ are the mid points of $B_{1}B_{3}and B_{2}B_{3}$ respectively. From the conjecture which states that the perpendicular bisector of any chord passes through the center of the circle, we can say that the intersection point of $L_{1} and L_{2}$ is at the center of the circle. Therefore, coordinates of the center of the Circle will be given as
\begin{equation}
\label{eqn_example}
X_c =  l * t_1 + \frac {X_1+X_3} {2}
\end{equation}
\begin{equation}
\label{eqn_example}
Y_c =  m * t_1 + \frac {Y_1+Y_3} {2}
\end{equation}
\begin{equation}
\label{eqn_example}
Z_c =  n * t_1 + \frac {Z_1+Z_3} {2}
\end{equation}
\begin{figure*}
    \centering
    \includegraphics[width=5in,height=3in]{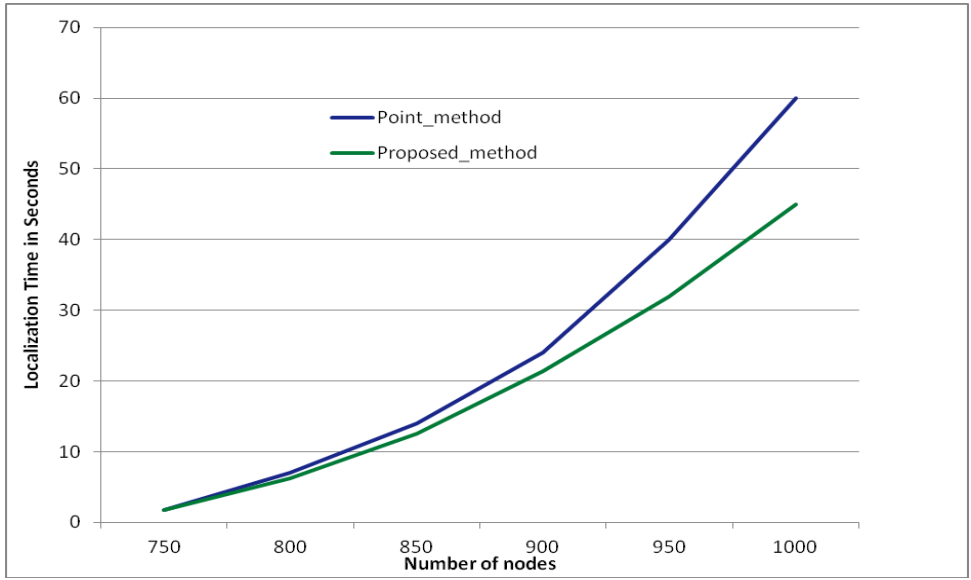}
    \caption{Localization Time Vs  \# of Nodes Localized}
    \label{LocalizationTime}
\end{figure*}
where $t_{1}$ is given as
\begin{equation}
\label{eqn_example}
t_1 = \frac {p(\frac{Y_1+Y_3}{2}-\frac{Y_3+Y_2}{2})-q(\frac{X_1+X_3}{2}-\frac{X_3+X_2}{2})}{q*l-p*m}
\end{equation}	
\textit{Equation of the line $L_{1}$:} passing through the center of the circle $C(X_{c}, Y_{c}, Z_{c})$ and perpendicular to the circle plane i.e. in the direction of  $\overrightarrow{N}$
\begin{equation}
\label{EquationOfLineL1}
\frac {X-X_c}{N_x}=\frac {Y-Y_c}{N_y}=\frac {Z-Z_c}{N_z}=t
\end{equation}
and we know that the line passing through center of circular cross section of the sphere and perpendicular to the plane of cross section will also pass through the center of the sphere.
Thus the center of sphere $S(X_{s}, Y_{s}, Z_{s})$ will satisfy the equation of line $L_{1}$ which gives us
\begin{equation}
\label{SphereCenterLieOnL1}
\frac {X_s-X_c}{N_x}=\frac {Y_s-Y_c}{N_y}=\frac {Z_s-Z_c}{N_z}=t
\end{equation}
On solving Equations (\ref{FirstRequiredEquation}), (\ref{SecondRequiredEquation}) and (\ref{SphereCenterLieOnL1}), we get the coordinates of the center of the sphere S i.e. $(X_{s}, Y_{s}, Z_{s})$ which is the required location of the sensor node and given as below:
\begin{equation}
\label{eqn_example}
X_s = t* N_x + x_c
\end{equation}
\begin{equation}
\label{eqn_example}
Y_s = t* N_y + y_c
\end{equation}
\begin{equation}
\label{eqn_example}
Z_s = t* N_z + z_c
\end{equation}%
where t can be computed from the Equation(\ref{ComputationForT})
\begin{figure*}[htp]
    \centering
    \includegraphics[width=5in,height=3in]{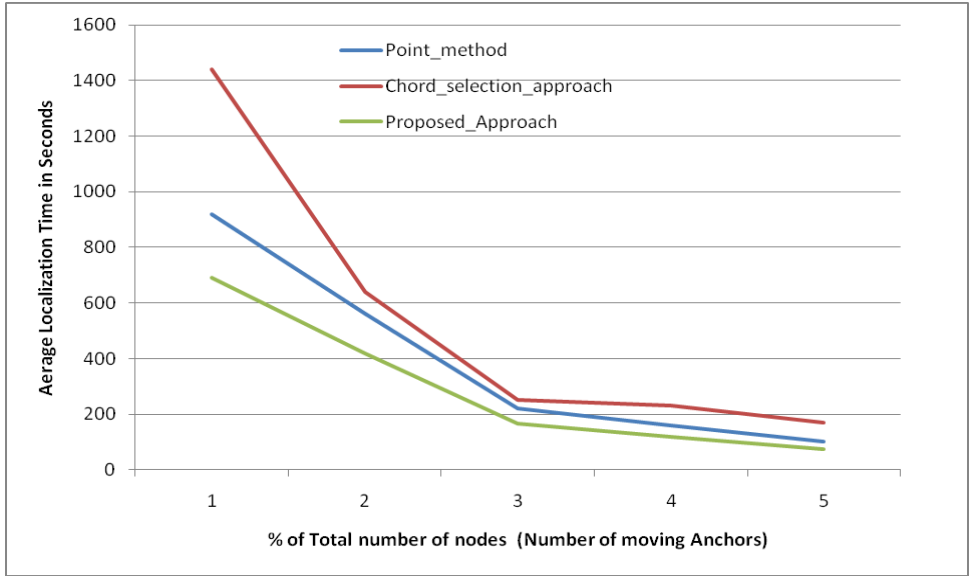}
    \caption{Average Localization Time Vs \# of Flying Anchors deployed for the process (\% of \# of Static sensors)}
    \label{AverageLocalizationTimeMovingAnchors}
\end{figure*}
\begin{figure*}[htp]
    \centering
    \includegraphics[width=5in,height=3in]{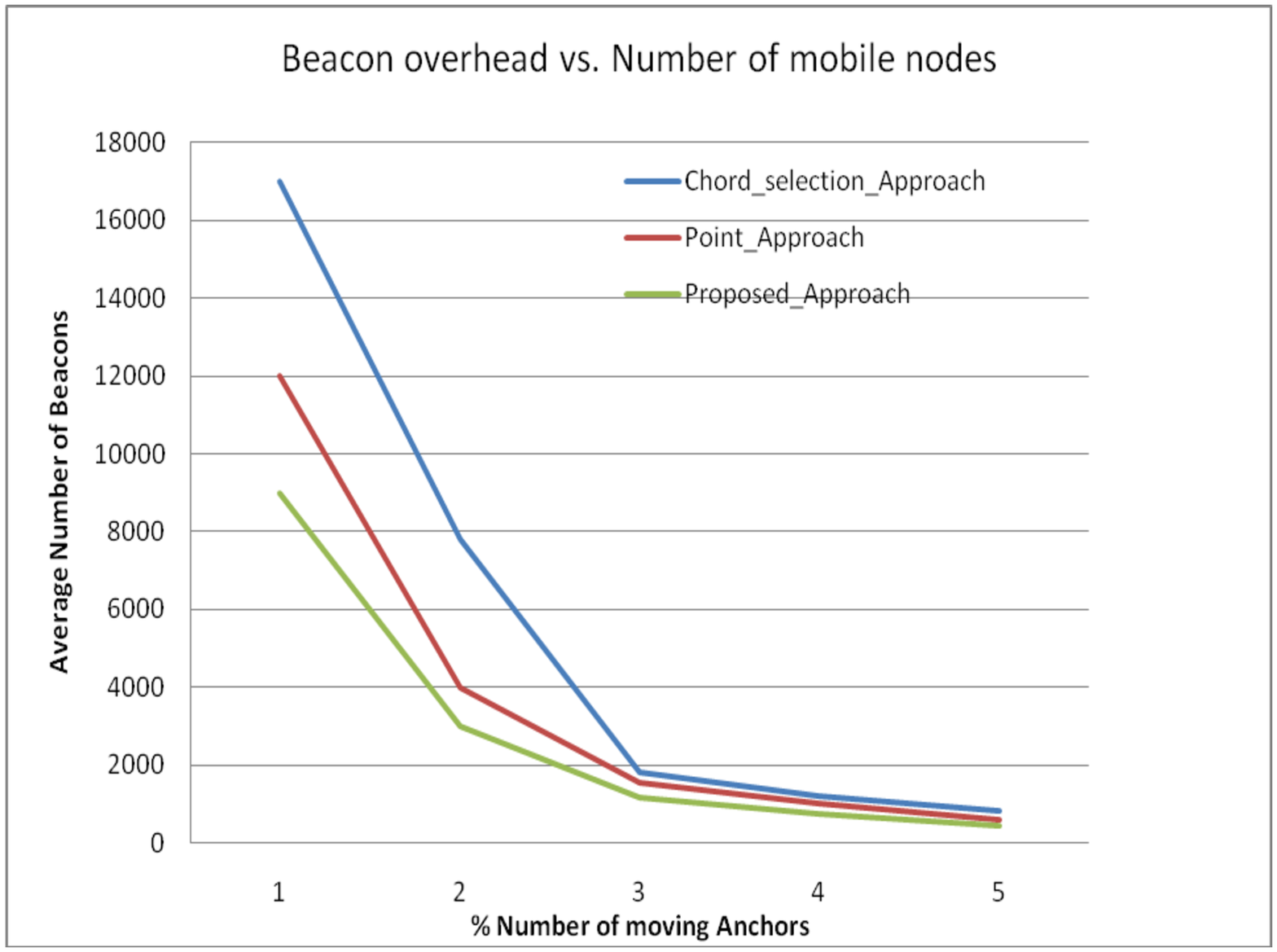}
    \caption{Average  \# of beacons required to localize Vs  \# of Flying anchors deployed for the process ((\% of \# of Static sensors))}
    \label{BeaconOverheads}
\end{figure*}
\section {Simulations and Results}
We have performed a series of simulation using Simulator for Networking Algorithm (SINALGO) which provides simulation framework for three dimensional networking algorithms. Simulations are performed on a region having volume of 1000 $\times$ 1000 $\times$ 1000 m\textsuperscript{3} and deployed 3000 static sensor nodes. The number of anchors used are 1\%, 2\%, 3\%, 4\%, and 5\% of total number of deployed static sensor nodes.\\
Anchor nodes moves according to random way point method and random direction walk. In random way point technique, the mobile nodes are deployed randomly. They randomly choose any destination and move in that direction with constant speed. Once they reach the destination they repeat the whole process. Random direction walk is similar to random way point walk. The only difference is the choice of the target. Instead of picking a random point from the deployment field, the random direction chooses a direction in which the node should walk, and how long the node should walk in this direction. If the node hits the boundary of the deployment area, it gets reflected back.
\subsection{Performance Metrics for Localization Algorithms }
Following performance metrics are generally used by researchers to evaluate any localization scheme which use similar approach to ours range free and based on to solve the problem of localization\cite{ChinesePaper}. \\
\textit{Average location error :} This metric is the average distance between the estimated location $(X_{e}, Y_{e}, Z_{e})$ and the actual location. $(X_{i}, Y_{i}, Z_{i})$ of all the sensor nodes, i.e.
\emph{Average location error}
\begin{equation*}
\label{AverageLocationError}
Error_{ALE}=\frac{\sum\sqrt{(X_{e}-X_{i})^2+(Y_{e}-Y_{i})^2+(Z_{e}-Z_{i})^2}}{Total \medspace  \#\medspace of \medspace Sensor \medspace Nodes}
\end{equation*}
\textit{Average localization time:} The average time taken for all the sensor nodes to compute their locations. It is given as \emph{average localization time}
\begin{equation*}
\label{AverageLocationTime}
Time_{ALT}=\frac{\sum{Localization \medspace time \medspace for \medspace each \medspace nodes }}{Total \medspace \# \medspace of \medspace Sensor \medspace Nodes}
\end{equation*}
\textit{Beacon Overhead: } The average  number of beacon messages broadcast by the moving anchors during the total localization time and given as
\emph{beacon overhead}
\begin{equation*}
\label{AverageLocationTime}
Overhead_{BO}=\frac{Total\medspace \#\medspace of\medspace beacon\medspace messages}{Total\medspace  \#\medspace of\medspace moving\medspace anchors}
\end{equation*}
In any sensor node most of the energy is consumed in computation and in message transmission. Therefore above two metrics defines energy efficiency.
\subsection{Analysis of Results}
We simulated our approach along with the methods proposed in \cite{MNNITPaper} and \cite{ChinesePaper} and observed that our approach outperforms both the approaches (\cite{MNNITPaper} and \cite{ChinesePaper}) in terms of average localization time, beacon overhead, number of moving sensors to be deployed and as well as communication overhead. The fig.\ref{LocalizationTime} demonstrates that proposed method performs significantly better than method proposed in \cite{MNNITPaper} (say. Point\_method).\\
Similarly, fig.\ref{AverageLocalizationTimeMovingAnchors} demonstrates the comparison of the performances of which were proposed in \cite{MNNITPaper} , \cite{ChinesePaper} and our proposed method. We found that our proposed method reduces average localization time significantly and is better than Point\_method and Chord\_selection\_approach.\\ Further, from the results shown in the fig.\ref{BeaconOverheads}, we see that average number of beacons required to localize the whole network is reduced significantly in compare to Point\_method and \\Chord\_selection\_approach.\\
Finally, \textit{average localization error} which has been discussed in detail in \cite{ChinesePaper} is the same in our approach. We did not simulate for all scenarios mentioned in \cite{ChinesePaper}. All other parameters such as chord selection (length), circular cross section selection, radio range and moving anchor velocity mentioned in \cite{ChinesePaper} which potentially affects the accuracy of the localization, will have the similar effect in our approach.
\section{Conclusion \& Future Work}
This paper proposes an efficient range-free localization scheme for three dimensional wireless sensor networks. Similar approaches use at least three known locations to localize one node while in our proposed approach we are using only three known locations without compromising the accuracy. The ability of localizing using only three known locations not only saves computations, time, energy, but also reduces the number of anchors need to be deployed and more importantly reduces the communication overheads.
The results show the significant improvement in every performance metric mentioned above. The performance of the proposed localization scheme is evaluated in a series of simulations have also been performed using SINALGO software which support three dimensional simulation. In future, utilizing the already localized nodes' locations in this approach will be interesting and should only improve the performance of the overall localizations process further.

\bibliography{references3}
\end{document}